\documentclass[10pt,twocolumn]{article}

\usepackage[margin=0.75in]{geometry}

\usepackage[utf8]{inputenc}
\usepackage[T1]{fontenc}
\usepackage{mathptmx} 
\usepackage{microtype}

\usepackage[dvipsnames]{xcolor}

\usepackage{booktabs}
\usepackage{array}

\usepackage[colorlinks=true,linkcolor=blue,citecolor=blue,urlcolor=blue]{hyperref}
\usepackage[numbers,sort&compress]{natbib}

\usepackage{graphicx}
\usepackage{float}
\usepackage{caption}
\usepackage{stfloats}

\definecolor{boxblue}{RGB}{230,242,255}
\definecolor{borderblue}{RGB}{0,102,204}

\newcommand{\keypoint}[1]{%
  \vspace{0.5em}%
  \noindent\fcolorbox{borderblue}{boxblue}{%
    \parbox{\dimexpr\columnwidth-2\fboxsep-2\fboxrule}{%
      \small #1%
    }%
  }%
  \vspace{0.5em}%
}

\newcommand{\keyquote}[1]{%
  \vspace{0.5em}%
  \noindent\fcolorbox{borderblue}{boxblue}{%
    \parbox{\dimexpr\columnwidth-2\fboxsep-2\fboxrule}{%
      \small\itshape #1%
    }%
  }%
  \vspace{0.5em}%
}

\makeatletter
\renewcommand{\section}{\@startsection{section}{1}{0pt}{-1.5ex plus -0.5ex minus -0.2ex}{0.8ex plus 0.2ex}{\large\bfseries}}
\renewcommand{\subsection}{\@startsection{subsection}{2}{0pt}{-1.2ex plus -0.4ex minus -0.2ex}{0.6ex plus 0.2ex}{\normalsize\bfseries}}
\makeatother

\title{\textbf{From ``Everything is a File'' to ``Files Are All You Need''}\\\large How Unix Philosophy Informs the Design of Agentic AI Systems}

\author{
  Deepak Babu Piskala\\
  \textit{Seattle, WA, USA}
}

\date{}

\begin{document}

\maketitle

\begin{figure*}[b!]
\centering
\includegraphics[width=0.75\textwidth]{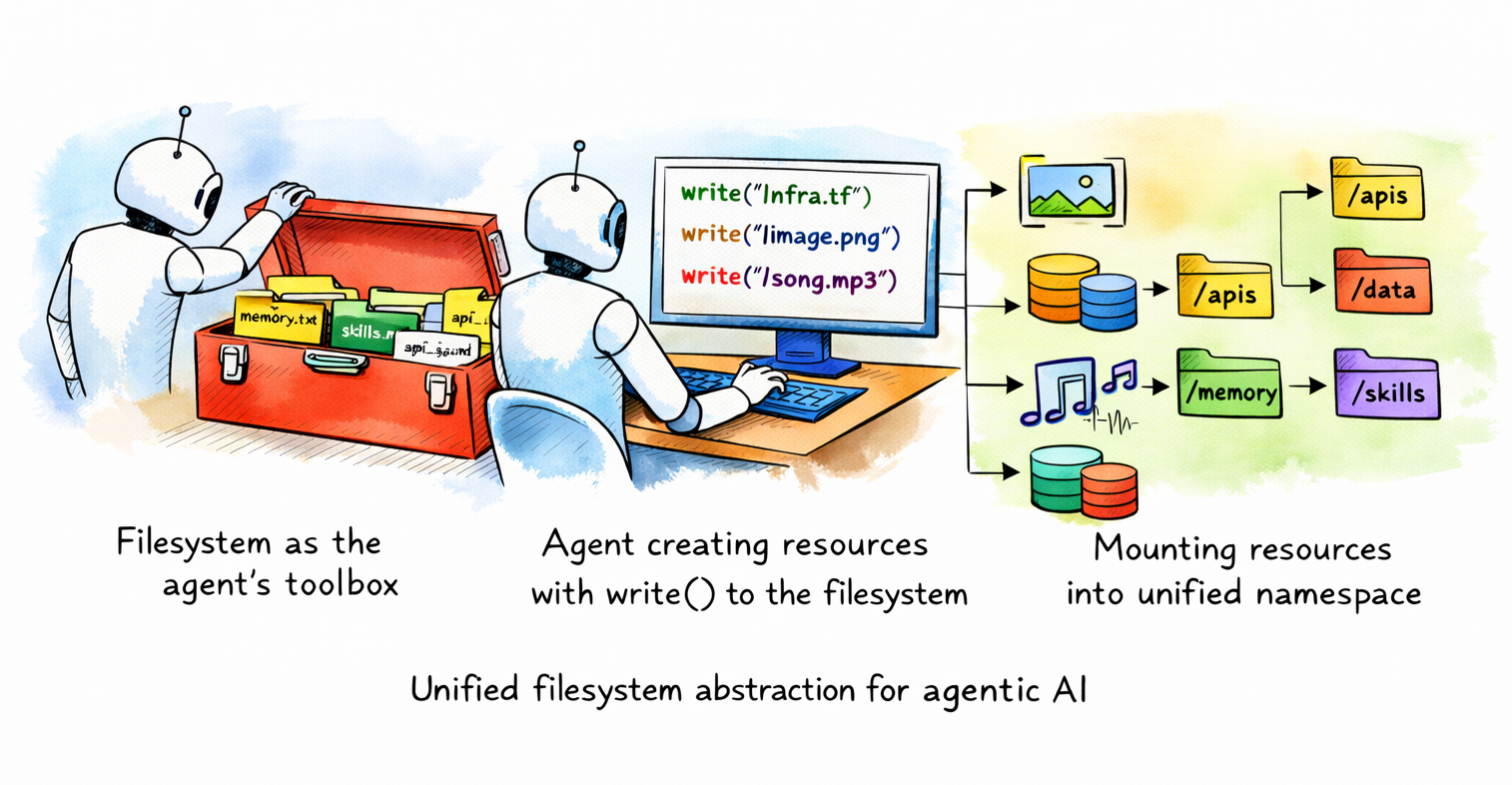}
\caption{Evolution of the uniform abstraction principle: from Unix's file system (1970s) through DevOps code artifacts (2010s) to agentic AI's file-based context and memory (2020s).}
\label{fig:evolution}
\end{figure*}

\begin{abstract}
\noindent This paper traces how a design philosophy from 1970s operating systems---Unix's ``everything is a file''---is finding new relevance in modern agentic AI. We examine the evolution from Unix through the DevOps movement to present-day autonomous agents, showing how the same principle of collapsing diverse resources into uniform interfaces continues to solve similar problems at different scales. For practitioners, the implication is clear: designing agents around file-like abstractions and code as the action language may offer a path to more maintainable, auditable systems.
\end{abstract}

\section{Introduction}

In the early 1970s, Ken Thompson and Dennis Ritchie faced a practical problem at Bell Labs. The computing systems of the day required programmers to learn different interfaces for different resources---one set of commands for disk access, another for terminal control, yet another for inter-process communication. The proliferation of special-purpose interfaces made systems difficult to learn, compose, and maintain.

Their solution was elegant in its simplicity: represent everything---devices, processes, network connections, data---as files. A hard drive and a terminal would both respond to the same operations: \texttt{open}, \texttt{read}, \texttt{write}, \texttt{close}. This wasn't the only possible design, but it proved remarkably durable. The abstraction survived decades of hardware evolution and remains foundational in Linux, macOS, and countless embedded systems today~\cite{ritchie1974unix}.

Half a century later, builders of autonomous AI systems face an analogous challenge. An agent that needs to query databases, call APIs, manage cloud infrastructure, store memories, and interact with humans confronts a similar proliferation of interfaces. Each capability comes with its own protocol, its own authentication scheme, its own data format. Without some unifying principle, the complexity threatens to become unmanageable.

What we're observing now is a convergence toward solutions that echo Unix's original insight. Coding agents like Claude Code and Cursor\footnote{Cursor: \url{https://cursor.sh}; Claude Code: \url{https://docs.anthropic.com/claude-code}} are increasingly treating filesystems as their primary substrate for context, memory, and action. Academic frameworks are proposing file-system abstractions for agent architectures~\cite{xu2025everything}. The pattern suggests that the old lesson---uniform interfaces reduce complexity---applies as much to AI agents as it did to operating systems.

This paper examines that convergence, tracing the thread from Unix through the DevOps era's ``everything as code'' movement to today's agentic AI designs. The goal is not to claim that files solve all problems, but to understand why this particular abstraction keeps proving useful and what that might mean for practitioners building agent systems today.

\section{The Unix Foundation}

The Unix philosophy emerged from constraints. Early computing systems were expensive and complex, and the engineers at Bell Labs sought ways to make them more tractable. One key insight was that simplifying the interface to resources---even at the cost of some flexibility---could yield large gains in usability and composability~\cite{wiki_eiaf}.

The file abstraction accomplished this by providing a universal handle. Whether a programmer wanted to read data from a disk, receive input from a keyboard, or communicate with another process, they used the same set of system calls. The underlying complexity didn't disappear, but it was encapsulated behind a consistent interface.

\keypoint{\textbf{The Unix Principle:} Expose diverse resources through a uniform interface. Complexity is encapsulated, not eliminated.}

This design enabled composition. Because programs read from and wrote to file descriptors using the same conventions, their outputs could feed directly into other programs' inputs. The Unix pipe---allowing commands like \texttt{ps aux | grep python | less}---became possible because all three programs spoke the same language of byte streams. Small, focused tools could be combined to accomplish complex tasks without any of them needing to know about the others.

The abstraction also lowered the barrier to understanding the system. A developer who learned how files worked could apply that knowledge broadly. New devices and virtual filesystems could be added without changing the fundamental model. This extensibility helps explain why Unix derivatives remain prevalent: the core abstraction continues to accommodate new requirements.

It's worth noting what the file abstraction doesn't do: it doesn't eliminate complexity, and it doesn't fit every use case perfectly. Device-specific control still sometimes requires escape hatches like \texttt{ioctl()}. But as a default model---a way of thinking about resources that covers most cases---the file metaphor has proven remarkably resilient.

\section{The DevOps Era: Extending the Pattern}

Beginning in the 2010s, the software industry began applying a similar principle to operations. The movement that became known as DevOps, and later GitOps, treated infrastructure, configuration, policies, and workflows not as ad-hoc settings to be clicked through in consoles, but as code to be versioned, reviewed, and tested.

The parallels to Unix are instructive. Just as Unix collapsed diverse device interfaces into file operations, the DevOps movement collapsed diverse operational concerns into code artifacts. The paradigm known as Infrastructure as Code (IaC)\footnote{IaC (sometimes abbreviated IaaC) treats infrastructure provisioning as software: servers, networks, and databases are defined in declarative configuration files rather than manually configured.}---moved infrastructure definitions from manual console configuration to declarative files in Terraform or CloudFormation. Deployment workflows moved from runbooks to YAML configurations in GitHub Actions or GitLab CI. Security policies moved from human-interpreted guidelines to machine-evaluable rules in Open Policy Agent or Sentinel.

Organizations that adopted these practices reported similar benefits to what Unix had provided: composability (pipelines could chain together), reproducibility (environments could be recreated from code), and auditability (version control tracked every change)~\cite{spotify_gitops}. HashiCorp articulated the principle explicitly:

\keyquote{``Everything has to be code---whether it's security, policy, automation, compliance, infrastructure---for all of the benefits we talked about.'' --- Seth Vargo, HashiCorp~\cite{hashicorp_vargo}}

The pattern extends well beyond infrastructure. Diagrams that once required manual drawing tools now emerge from text descriptions: Mermaid.js\footnote{Mermaid is a JavaScript-based tool that renders diagrams from markdown-like text definitions, enabling version-controlled documentation.} and PlantUML generate flowcharts, sequence diagrams, and architecture charts from markdown-like syntax~\cite{mermaid_docs}. API contracts move from informal documentation to machine-readable OpenAPI and GraphQL schemas~\cite{openapi_spec}. Database changes become versioned migration scripts. Data quality rules become executable specifications.

\begin{table}[h]
\centering
\small
\begin{tabular}{@{}p{1.8cm}p{2.2cm}p{2.5cm}@{}}
\toprule
\textbf{Domain} & \textbf{Traditional} & \textbf{Code-Based} \\
\midrule
Infrastructure & Console clicks & Terraform, Pulumi \\
Diagrams & Drawing tools & Mermaid, PlantUML \\
API specs & Documents & OpenAPI, GraphQL \\
DB changes & Manual SQL & Flyway, Alembic \\
Data quality & Manual checks & Great Expectations \\
Presentations & PowerPoint & Slidev, reveal.js \\
\bottomrule
\end{tabular}
\caption{The ``as code'' pattern across domains}
\label{tab:ascode}
\end{table}

What unites these examples is the same move Unix made with files: replace heterogeneous interfaces with text-based representations that can be versioned, diffed, tested, and composed.

\section{Agentic AI and the Return of the File}

Today's agentic AI systems---autonomous agents that plan, execute, and adapt---face the interface proliferation problem in acute form. A capable agent might need to interact with REST APIs, SQL databases, vector stores, cloud consoles, file systems, web browsers, and human users, each with different protocols and conventions.

Recent observations suggest that practitioners are converging on filesystems and code as the unifying abstraction. Jerry Liu documented three patterns emerging in coding agents~\cite{liu2026files}:

\textbf{Pattern 1: Conversation History Storage.} When context windows fill up and compaction occurs, agents lose their working memory. Storing conversations to searchable files provides a way to maintain continuity across sessions---the agent can read back what it previously discussed rather than starting fresh.

\textbf{Pattern 2: Context Retrieval via Files.} Rather than relying solely on vector similarity search, agents traverse file systems, interleaving search operations with reading specific files or sections. This mimics how a human might scan through documents, drilling down where relevant.

\textbf{Pattern 3: Skills Replacing Tools.} Anthropic's concept of ``skills''---essentially markdown files describing how to accomplish tasks---allows agents to learn capabilities by reading documentation rather than through programmatic tool definitions. Instead of defining a custom MCP\footnote{Model Context Protocol (MCP) is Anthropic's standard for connecting AI agents to external tools and data sources.} tool with programmatic code for each capability, a developer can drop an API specification or workflow description into a markdown file; the agent reads and follows it just as a human developer would read documentation.\footnote{Anthropic Skills documentation: \url{https://docs.anthropic.com/claude/docs/skills}} This is the file abstraction applied to agent capabilities themselves.

\keyquote{``Reasoning agents with file system tools and semantic search lets agents dynamically traverse context to answer questions of any complexity.'' --- Jerry Liu~\cite{liu2026files}}

\textbf{Multi-Agent Research Systems.} Anthropic's engineering team has documented how their multi-agent research system relies heavily on file-like memory abstractions~\cite{anthropic_research}. When a user submits a complex research query, a lead agent analyzes it, develops a strategy, and saves its plan to a Memory component to persist context---critical because context windows exceeding 200,000 tokens will be truncated. The lead agent then spawns specialized subagents that search in parallel, each with their own context windows. Subagents write findings back to shared memory, and the lead agent reads these results to synthesize a final answer. The entire architecture treats memory as a file-like resource: agents save plans, retrieve context, and pass findings through read/write operations on a shared namespace.

Academic work has begun to formalize this approach. The AIGNE framework proposes an ``Agentic File System'' where heterogeneous resources---memory stores, tools, external APIs---are mounted into a unified namespace~\cite{xu2025everything}. The framework treats context engineering as a matter of selecting, compressing, and loading files. History, memory, and scratchpads become file-like resources with metadata and access controls.

\keypoint{\textbf{The File System Advantage:} A familiar, composable interface that agents can manipulate using a small set of operations---\texttt{list}, \texttt{read}, \texttt{write}, \texttt{search}, \texttt{execute}.}

This isn't to claim that files solve every problem. Non-text formats require parsing. Massive document collections require indexing beyond simple grep. Some interactions don't map naturally to file operations. But as a default abstraction, the filesystem appears to be finding renewed utility in agentic AI.

\section{Code as the Action Language}

The filesystem provides structure for context; code provides a language for action. These two elements work together.

Coding models succeed in part because code offers properties that natural language lacks. A SQL query is unambiguous in a way that ``find the relevant customer records'' is not. A Python script can be tested, debugged, and sandboxed. A Terraform plan can be previewed before execution. Code makes it possible to verify, constrain, and audit what an agent does.

The practical implication is that agents built around code generation often require fewer custom tools. Rather than defining a specific tool for querying a database, the agent writes SQL. Rather than defining a tool for API calls, the agent writes a Python request or curl command. The agent becomes a ``code transformer''---receiving goals in natural language and producing code that accomplishes them.

\keyquote{``The agent really only has access to a filesystem and $\sim$5--10 tools: CLI over filesystem, code interpreter, web fetch---and this is just as general, if not more general, than agent with 100+ MCP tools.'' --- Jerry Liu~\cite{liu2026files}}

This parallels the Unix observation again. Unix reduced N device-specific APIs to a handful of system calls; coding agents reduce N service-specific tool definitions to code generation plus execution. The surface area shrinks, and what remains is more uniform.

\section{The Core Pattern}

What connects Unix, DevOps, and agentic AI is a recurring strategy for managing complexity:

\keypoint{\textbf{The Unifying Strategy:} Collapse diverse interfaces into a uniform abstraction, accepting some loss of specialization in exchange for composability and tractability.}

The specific abstraction changes---files, code, file-like resources---but the underlying move is the same. In each case, practitioners faced a proliferation of special-purpose interfaces that made systems hard to build, understand, and maintain. In each case, introducing a common abstraction reduced the cognitive and engineering burden, enabling composition and reuse.

This doesn't mean files or code are always the right answer. It means that when confronted with interface explosion, looking for a unifying abstraction is a strategy that has paid off repeatedly across computing history.

\section{Implications for Practitioners}

For engineers building agentic systems, several practical considerations follow:

\textbf{Consider file-like abstractions for resources.} Whether dealing with memory, tools, or external services, exposing them through a namespace that agents can traverse with standard operations may simplify agent design~\cite{xu2025everything}.

\textbf{Let agents generate code rather than accumulating tool integrations.} Code generation allows agents to interact with APIs, databases, and services without requiring dedicated integration code for each. The tradeoff is that code execution requires sandboxing, but these are well-understood problems.

\textbf{Leverage DevOps infrastructure.} Version control, CI/CD pipelines, GitOps workflows---the tools built for managing code artifacts apply equally well to agent artifacts.

\textbf{Maintain traceability.} Every substantive agent action should produce an auditable artifact: a diff, a log entry, a commit. File-based abstractions make this natural.

\textbf{Expect gaps.} Non-plaintext documents need parsing. Large collections need indexing. Some interactions don't fit file metaphors. These are known limitations, not reasons to reject the abstraction.

\section{Conclusion}

The file abstraction has proven useful for half a century because it addresses a recurring problem: how to manage systems that must interact with many different kinds of resources. The solution---providing a uniform interface that hides heterogeneity---reduces complexity at the cost of some specificity.

Agentic AI systems face the same fundamental challenge and are arriving at similar solutions. Files provide structure for context; code provides a language for action. Together, they offer a tractable substrate for agents that must navigate complex digital environments.

Whether this pattern will remain dominant is uncertain. New abstractions may emerge as the field develops. But for now, the convergence is notable: a design principle from early operating systems is finding renewed relevance in one of computing's newest frontiers.

\keypoint{\textbf{The Throughline:} Unix showed that when everything looks like a file, systems become more composable and maintainable. For agentic AI, when everything looks like code within a filesystem-like structure, similar benefits may follow.}

\bibliographystyle{unsrtnat}

\end{document}